\documentclass[pre,aps,twocolumn]{revtex4}
\usepackage{amsmath,bm,epsfig}
 \newcommand{\B}[1]{{\bm{#1}}}

\usepackage[latin1]{inputenc}
\def\BE{\begin{equation}}\def\EE{\end{equation}}
\def\BEA{\begin{eqnarray}}\def\EEA{\end{eqnarray}}
\def\BSE{\begin{subequations}}\def\ESE{\end{subequations}}
\def\<{\left\langle} \def\>{\right\rangle} \def\({\left(} \def\){\right)}
\let \= \equiv
  \let\~\widetilde \let\^\widehat 
   
  \def\1{\bm1}

\def\Fbox#1{\vskip1ex\hbox to 8.5cm{\hfil\fboxsep0.3cm\fbox{%
  \parbox{8.0cm}{#1}}\hfil}\vskip1ex\noindent}  %%  {TEXT} in BOX
\begin{document}
\title{Turbulent Drag Reduction by Flexible and Rodlike Polymers: Crossover
Effects at Small Concentrations}
\author{Emily S.C. Ching}
\affiliation{Dept. of Physics, The Chinese University of Hong
Kong, Shatin, Hong Kong}
\author{T. S.  Lo}
\affiliation{Dept. of Chemical Physics, The Weizmann Institute
of Science, Rehovot 76100, Israel}
\author{Itamar Procaccia}
\affiliation{Dept. of Chemical Physics, The Weizmann Institute
of Science, Rehovot 76100, Israel}
\begin{abstract}
Drag reduction by polymers is bounded between two universal asymptotes, the von-K\'arm\'an
log-law of the law and the Maximum Drag Reduction (MDR) asymptote.
It is  theoretically understood why the MDR asymptote is universal, independent
of whether the polymers are flexible or rodlike. The cross-over
behavior from the Newtonian von-K\'arm\'an log-law to the MDR is
however not universal, showing different characteristics for
flexible and rodlike polymers. In this paper we provide a theory
for this cross-over phenomenology.
\end{abstract}
\maketitle
\section{Introduction}
%%%%%%Figure 1%%%%%%%%%%
\begin{figure}
\hskip -1
 cm
\centering \epsfig{width=.50\textwidth,file=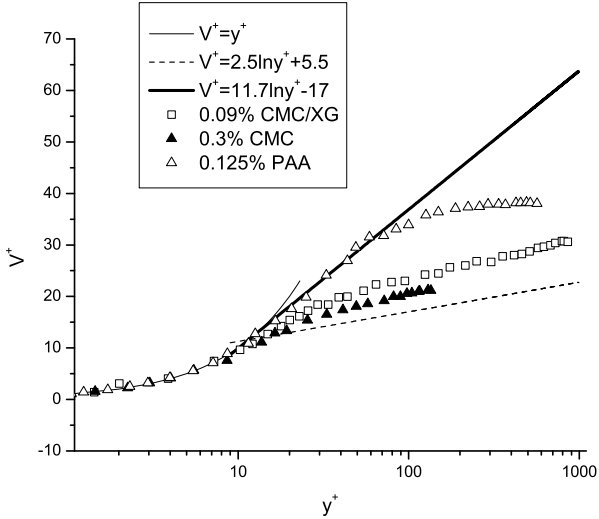}
\caption{Typical velocity profiles taken from \cite{Escudier}. In
dashed line we noted the von-K\'arm\'an law (\ref{LLW}), while the
MDR (\ref{final}) is shown as the continuous black line. In all
cases the mean velocity follows the same viscous behavior for
$y^+<10$. After that the scenario is different for flexible and
rodlike polymers. The typical behavior for the former is presented
by the open triangles, which follow the MDR up to a crossover
point that depends on the concentration of the polymer and on the
value of Re. The rodlike behavior is exemplified by the solid
triangles and the open squares; the mean velocity profiles appear
to interpolate smoothly between the two asymptotes as a function
of the concentration of the rodlike polymer.} \label{expt}
\end{figure}
%%%%%%%%%%%%%%%%%%

The phenomenon known as ``drag reduction" by polymers in turbulent
channel flows \cite{Book1,Virk} is conveniently discussed in
channel geometry for fixed pressure heads, such that the reduction
in the drag is manifested as an increased mean velocity. For the
sake of comparison between different fluids it is convenient to
choose normalized coordinates. Denote the mean pressure (per unit density) 
gradient $p'\equiv -\partial p/\partial x$ where $x$, $y$ and $z$ are the
lengthwise, wall-normal and spanwise directions respectively. The
length and width of the channel are usually taken much larger than
the mid-channel height $L$, making the latter a natural re-scaling
length for the introduction of dimensionless (similarity)
variables, also known as ``wall units" \cite{Pope}. Thus the
Reynolds number Re, the normalized distance from the wall $y^+$
and the normalized mean velocity $V^+(y^+)$ (which is in the $x$
direction with a dependence on $y$ only) are defined by
\begin{equation}
{\rm Re} \equiv {L\sqrt{\mathstrut p' L}}/{\nu_0}\ , \  y^+ \equiv
{y {\rm Re} }/{L} \ , \  V^+ \equiv {V}/{\sqrt{\mathstrut p'L}} \
, \label{red}
\end{equation}
where $\nu_0$ is the kinematic viscosity of the neat fluid. For
Newtonian fluids the profile of the mean velocity $V^+(y^+)$ is
universal, starting with the viscous sub-layer in which
$V^+(y^+)=y^+$ and then, at $y^+$ somewhere between 6 and 12, the
profile crosses over to the universal von-K\'arm\'an log-law of
the wall (cf. Fig. \ref{expt})
\begin{equation}
\label{LLW} V^+(y^+) =\kappa_{_{\rm K}}^{-1}\ln y^+ + B\ .
\end{equation}
Upon the addition of small concentrations of polymers, the drag is
reduced and for the same value of $p'$ one finds an increase in
$V^+(y^+)$. This phenomenon exhibits both universal and
non-universal aspects. The universal aspect is the Maximum Drag
Reduction (MDR) asymptote, which is the largest attainable profile
$V^+(y^+)$. This was determined experimentally by Virk who found
\cite{Virk}
\begin{equation}
V^+(y^+) = \frac{1}{\kappa_{_{\rm V}}}\ln\left(e\, \kappa_{_{\rm
V}} y^+\right)\, \quad{\rm for}~ y^+ \gtrsim 12
 \ . \label{final}
\end{equation}
While $\kappa_{_{\rm K}}^{-1}\approx 2.5$, $\kappa_{_{\rm
V}}^{-1}\approx 12$, leading to a significantly larger mean
velocity at the MDR as compared with von-K\'arm\'an's log-law. The
MDR appears independent of the nature of the polymer (for example
whether it is flexible or rodlike), of the length of the polymer
and of the concentration. On the other hand, the way that the
system {\em attains} the MDR is not universal, and it depends on
all of the above. In particular, it appears that flexible and rodlike
polymers attain the MDR, as a function of the concentration, in
qualitatively different ways (see for example \cite{Virk}). The
experimental information is quite scant, but available data
indicate different scenarios for the two types of polymers. The
data in Fig. \ref{expt} indicate for large values of Re the mean
velocity profile with flexible polymers [polyacrylamide (PAA)] 
follows the MDR until a
point of cross over back to the ``Newtonian plug",  where it
becomes parallel to von-K\'arm\'an's log-law. Increasing the
concentration results in following the MDR further until a higher
cross over point is attained back to the Newtonian plug
\cite{Virk}.  On the other hand, for rodlike polymers 
[sodium carboxymethylcellulose (CMC) and 
sodium carboxymethylcellulose/xanthan gum blend
(CMC/XG)] the data
shown in Fig. \ref{expt} indicate a different scenario. Contrary
to flexible polymers, here, as a function of the concentration,
one finds mean velocity profiles that interpolate between the two
asymptotes (\ref{LLW}) and (\ref{final}), reaching the MDR only
for large concentrations. A difference in the behavior of
flexible and rodlike polymers 
when plotting the drag as a function of Reynolds number
was also reported by Virk and coworkers, see \cite{97VSW}.

The universal MDR was fully explained in recent work, and its
parameters derived, by showing that it is an edge turbulent
solution in a channel \cite{master}. In other words, if one tried
to reduce the drag further (or further increase the profile
$V^+(y^+)$), one would lose the turbulent flow in favor of a
laminar solution. This is the reason for the universality of the
MDR and its insensitivity to the nature of the polymer. The aim of
this paper is to derive quantitatively the non-universal scenarios
of attaining the MDR by flexible and rodlike polymers. We will
limit our attention to the case of high Re, and consider the
profiles $V^+(y^+)$ for different concentrations, with the aim of
explaining the phenomenology displayed in Fig.  \ref{expt}. The
paper ends with a definite prediction of different crossover
behavior in the mean velocity profiles for flexible and rodlike
polymers, in agreement with the indication of Fig. \ref{expt}.

In Sect. 2 we summarize the available theory of drag reduction by
flexible and rodlike polymers, based on the balance equations for
energy and momentum in the turbulent boundary layer. We derive in
this section the equations satisfied by the $y$ dependence of the
mean shear (in wall units), from which the velocity profiles are
obtained by integration. Sect. 3 presents the results together
with a summary and a discussion.
%%%%%%%%%%%%%%%%%%%%%%%%%%%%%%%%%%%%%
\section{Theory of Drag Reduction}
\subsection{The polymeric stress tensor}
In the presence of a small concentration of polymers the Navier-Stokes equations for the
fluid velocity $\B U(\B r,t)$ gain an additional stress tensor:
\begin{eqnarray}
\label{Navier}
 \frac{\partial \B U}{\partial t} +\B U\cdot \B \nabla \B U&=& - \nabla  p +\nu_0 \nabla^2
\B U  +\nabla \cdot {\B \sigma} \ , \\
\nabla \cdot \B U &=& 0 \ . \nonumber
\end{eqnarray}
The extra stress
tensor $\B \sigma$ is due to  the interaction between the polymers
and the fluid.  Within the FENE-P model for flexible polymers we have \cite{JFM}
\begin{equation}
\label{flex} \mathcal{\sigma}_{ab} \approx \nu_p \gamma_p
\mathcal{R}_{ab} \ ,
\end{equation}
where $\nu_p$ is the polymeric contribution to the viscosity at
zero shear, and $\gamma_p$ is the inverse relaxation time of the
stretched polymer. In writing this expression one adopts the
standard simplification of a single relaxation time. The
conformation tensor $\mathcal{ \B R }$ is obtained from the
normalized end-to-end distance vector of the polymer $ \hat {\B
\rho }\equiv \B \rho/\rho_{\rm max}$, averaged over the
conformation of the polymers,
\begin{equation}
\B{\mathcal{ R }} \equiv \overline{\hat {\B \rho }\hat { \B \rho} }
\end{equation}

For rodlike polymer $ \hat {\B  \rho }$ becomes a unit vector, and
there is no coil-stretching transition. Accordingly  the stress
tensor assumes a different form \cite{88DE}:
\begin{equation}
\label{rigid} \sigma_{ab}=6 \nu_p \mathcal{R}_{ab}
\mathcal{R}_{ij} S_{ij} \ ,
\end{equation}
where $S_{ij}$ is the strain experienced by the polymer $S_{ij}
\equiv \partial U_i / \partial r_j $. As explained above, the
difference in form of the stress tensor is immaterial for the
universal form of the MDR \cite{PRE,Virk}, a  phenomenon that was
called  ``additive equivalence'' by Virk. We will see that this
difference translates however to a very different scenario for the
attainment of the MDR.
%%%%%%%%%%%%%%%%%
\subsection{The balance equations}
The phenomenon of drag reduction can be understood
on the basis of the balance equations of the mechanical momentum and turbulent
energy \cite{PRL}.  These are derived on the basis of the Reynolds decomposition
\begin{eqnarray}
U_i(\B r,t) &= &V(y)\delta_{ix} + u_i(\B r, t)\\
S_{ij} (\B r,t)&=& S(y) \delta_{ix}\delta_{jy} +s_{ij}(\B r,t) \ ,
\quad S(y) \equiv \frac{dV(y)}{dy} \ .
\end{eqnarray}
Writing these equations we use the fact that in a turbulent
channel flow $p'$ is constant, and due to the symmetry all the
other mean quantities depend on $y$ only. In addition to the mean
shear $S(y)$, we need to to introduce the mean turbulent kinetic
energy $K \equiv \langle u^2 \rangle/2$ and the Reynolds stress $W
\equiv - \langle u_x u_y \rangle$. The momentum balance equation
is obtained by averaging Eq. (\ref{Navier}) and integrating with
respect to $y$, ending up with the exact equation:
\begin{equation}
\label{raw_mom} \nu_0 S + W + \langle \sigma_{xy} \rangle = p' (L-y) \ .
\end{equation}

In wall-units Eq. (\ref{raw_mom}) can be written in a more elegant form:
\begin{equation}
\label{mom} S^+ + W^+ +\langle \sigma^+_{xy} \rangle  = (1-y^+
/{\rm Re})
\end{equation}
where $S^+ \equiv \nu_0 S /(p'L)$,  $W^+ \equiv W/(p'L)$, and
$\sigma_{ij}^+ \equiv \sigma_{ij}/(p'L)$.
When Re is very large and for $y^+$ not too large we neglect
the second term on the RHS, approximating the RHS as unity.

The balance equation for the turbulent kinetic energy is calculated by taking
the dot product of the fluctuation part of Eq. (\ref{Navier}) with $\B u$:
\begin{equation}
\label{raw_energy} W  S  = \frac{\partial } {\partial y } \langle
u_y u^2  +  u_y p  - \sigma_{iy}u_i  \rangle  + \nu_0 \langle
s_{ij} s_{ij} \rangle \ + \langle \sigma_{ij} s_{ij} \rangle .
\end{equation}
Also this equation is exact. We simplify it by noting that
the first term on the  RHS involves the spatial flux of turbulent energy
 which is known to be negligible in the log-layer.  The second term
represents the dissipation which is modelled (in wall-units) as
\cite{PRL}:
\begin{equation}
\langle s^+_{ij} s^+_{ij} \rangle \ \approx K^+ (\frac{a}{y^+})^2
+ b\frac{(K^+)^{3/2}}{y^+} \ .
\end{equation}
where $s_{ij}^+ \equiv \nu_0 s_{ij}/(p'L)$ and $K^+ \equiv K/(p'L)$.
Therefore, the energy balance equation takes on the form
\begin{equation}
\label {energy} W^+ S^+ = K^+ (\frac{a}{y^+})^2 +
b\frac{(K^+)^{3/2}}{y^+} + \langle \sigma^+_{ij} s^+_{ij} \rangle
\end{equation}
Finally, we quote the experimental fact that in the log-layer
$W^+$ and $K^+$ are proportional to each other:
\begin{equation}
\label{KW} K^+ c^2 = W^+ \ .
\end{equation}
Experimentally, it was found that $c \approx 0.5$ in the Newtonian
case, and $c \approx 0.25$ in the MDR.

\subsection{Effect of the polymers}
In the Newtonian case, when $\B \sigma=0$ and $c=0.5$, the three
equations (\ref{mom}), (\ref{energy}) and (\ref{KW}) are
sufficient for determining the three unknowns $S^+$, $K^+$ and
$W^+$. The best fit to Newtonian experiments and simulations are
obtained using the values  $a=3$ and $b=0.321$, and we are going
to use these values throughout, also in the viscoelastic cases
discussed below. In the presence of polymers, however, we have to
consider the terms introduced by the polymers to these equations.
The necessary theory was presented in  \cite{PRE,PRL}, with the
final results relating these terms to the $yy$ component of the
mean conformation tensor $\B R\equiv \langle \B{\mathcal{
R}}\rangle$: .
\begin{eqnarray}
\langle \sigma_{xy}\rangle & \approx &  {\nu_p} R_{yy} S \ , \\
\langle \sigma_{ij}s_{ij} \rangle & \approx & {\nu_p} R_{yy}
\frac{K}{y^2} \  . \nonumber
\end{eqnarray}
Using these theoretical results the momentum and energy equations, in the
limit of large Re, are reduced to
\begin{equation}
\label{mom1} S^+ + W^+ + \tilde{\nu} R_{yy} S^+ = 1 \ ,
\end{equation}
and
\begin{equation}
\label{en1} W^+ S^+ = K^+ (\frac{a}{y^+})^2 +
b\frac{(K^+)^{3/2}}{y^+} + \tilde{\nu} R_{yy} K^+ (\frac{a}{y^+})^2
\  .
\end{equation}
Here $\tilde \nu = \nu_p/\nu_0$. To proceed we need to relate $R_{yy}$ to the
other variables. The necessary theory is presented in  \cite{paperB, PRE,PRL,JFM}
leading to an important difference  between the flexible and the
rodlike polymers:
\begin{eqnarray}
R_{yy} &\approx& \frac{\sqrt{K}}{Sy} \qquad (\mbox{flexible}) \ ,  \\
R_{yy} &\approx& \frac{K}{S^2 y^2} \qquad (\mbox{rodlike})
\end{eqnarray}
Substituting into  Eqs.~(\ref{mom1}) and (\ref{en1}) we have
\begin{equation}
\nu_{\rm eff} S^+ + W^+  = 1\ ,
\label{eq1}
\end{equation}
and
\begin{equation}
W^+ S^+ = K^+ \nu_{\rm eff} (\frac{a}{y^+})^2 + b\frac{(K^+)^{3/2}}{y^+} \ ,
\label{eq2}
\end{equation}
with the ``effective viscosity"
\begin{eqnarray}
\nu_{\rm eff} &=& 1 + \tilde{\nu} \frac{\sqrt{K^+}}{S^+y^+}
\qquad (\mbox{flexible}) \ ,  \label{nu1}\\
\nu_{\rm eff} &=& 1 + \tilde{\nu} \frac{K^+}{(S^+ y^+)^2} \qquad (\mbox{rodlike}) \ .
\label{nu2}
\end{eqnarray}

In the next section we will show that although the effective viscosities take on different
forms for the flexible and rodlike cases, in fact they both depend linearly on $y^+$ whenever
we have a log layer with $S^+\propto 1/y^+$. The reason is that $K^+$ turns out to be
proportional to $(y^+)^2$ and to $y^+$ for the flexible and the rodlike cases respectively.
Accordingly, we will write in both cases
\begin{equation}
\nu_{\rm eff}= 1 + \alpha[y^+-\Delta(\alpha)] \ , \label{nueff}
\end{equation}
with $\Delta(\alpha)$ being the width of the viscous sub-layer, and its dependence
on the slope of the effective viscosity $\alpha$
needs to be determined. It is natural to present $\Delta(\alpha)$  in terms of
a dimensionless scaling function $f(x)$,
\begin{equation}
\Delta(\alpha) =\delta^+ f(\alpha\delta^+) \ , \label{scaling}
\end{equation}
where $\delta^+\approx 6$ is the width of the Newtonian viscous
boundary layer. In the Newtonian limit $\alpha \to 0$, $\nu_{\rm eff} \to 1$ and $\Delta \to \delta^+$,
hence we have $f(0)=1$. In \cite{master} it was shown
that the balance equations (\ref{eq1}) and (\ref{eq2}) (with the
prescribed form of the effective viscosity profile) have an
non-trivial symmetry  that leaves them invariant under rescaling
of the wall units.  This symmetry dictates the function
$\Delta(\alpha)$ in the form
\begin{equation}
\Delta(\alpha) =\frac{\delta^+}{1-\alpha\delta^+} \  . \label{Delta}
\end{equation}
%%%%%%%%%%%%%%%%%%%%%%%%%%
\subsection{Closing the equations}

To complete the model, we have to specify the value of $c$ in Eq.
(\ref{KW}).  This parameter becomes naturally a function of
$\alpha$. We can find its $\alpha$-dependence by identifying the
width of the viscous sub-layer $\Delta$ with $a/c(\alpha)$. This
stems from the fact that the balance equations cannot support a
turbulent solution for $y^+<a/c(\alpha)$. This means that
\begin{equation}
a/c(\alpha)=\Delta(\alpha) \ . \label{a/c}
\end{equation}
Combining then Eqs. (\ref{nueff}), (\ref{Delta}) and (\ref{a/c}),
and putting $\delta^+=6$,  we can solve and find
\begin{eqnarray}
\nonumber
c(\nu_{\rm eff})
&=& \frac{a}{6} + \frac{a(2-\nu_{\rm eff})}{2y^+} \\
&+& \frac{a}{12} \left[
\sqrt{1-\frac{12 \nu_{\rm eff}}{y^+}
+\frac{36(2-\nu_{\rm eff})^2}{(y^+)^2}} - 1 \right]
\end{eqnarray}

To summarize note that
Eqs.~(\ref{nu1}) and (\ref{nu2}) can be written as
\begin{equation}
K^+ = A^2 (S^+y^+)^2
\label{K}
\end{equation}
with
\begin{eqnarray}
A^2 &=& \left(\frac{\nu_{\rm eff} -1 }{\tilde \nu}\right)^2 \qquad (\rm{flexible}) \\
A^2 &=& \frac{\nu_{\rm eff} -1 }{\tilde \nu} \qquad \qquad (\rm{rodlike})
\end{eqnarray}

Using Eqs.~(\ref{KW}) and (\ref{K}), we can rewrite Eqs.~(\ref{eq1}) and (\ref{eq2})
as two equations for the two
variables $\nu_{\rm eff}$ and $S^+$:
\begin{equation}
\nu_{\rm eff} S^+ + c^2 A^2 (S^+ y^+)^2  = 1 \ ,
\label{Neq1}
\end{equation}
and
\begin{equation}
c^2 S^+ = \nu_{\rm eff} (\frac{a}{y^+})^2
+ b A S^+ \
\label{Neq2}
\end{equation}

Equation (\ref{Neq2}) implies
\begin{equation}
S^+ = \frac{\nu_{\rm eff}}{(y^+)^2} \ \frac{a^2}{(c^2-bA)} \label{S}
\end{equation}
Substituting Eq.~(\ref{S}) into Eq.~(\ref{Neq1}) gives an equation
for $\nu_{\rm eff}$:
\begin{equation}
\nu_{\rm eff}^2 (\frac{a}{y^+})^2 (c^2-bA) + c^2A^2 \nu_{\rm
eff}^2 (\frac{a^2}{y^+})^2 = (c^2-bA)^2 \label{Fnueff}
\end{equation}

Finally, we can solve
Eq. (\ref{Fnueff}) to get $\nu_{\rm eff}(y^+)$ for different values of $\tilde \nu$.
Then we can obtain $S^+$ and $K^+$ using Eqs.~(\ref{S}) and (\ref{K}) respectively.
Integrating $S^+$ over $y^+$, we get also $V^+(y^+)$.

%%%%%%%%%%%%%%%%%%%%%%%%%%%%%%%
\section{Results}

\begin{figure}
\centering \epsfig{width=.40\textwidth,angle=-90,file=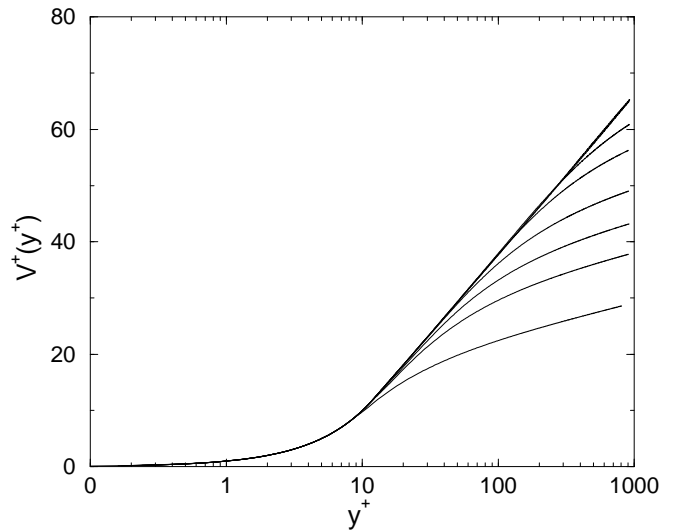}
\caption{The mean velocity profiles for flexible polymer additives
with $\tilde \nu=1,5,10,20,50, 100$ and 500 from below to above.
Note that the profile follows the MDR until it crosses over
back to the Newtonian plug.}
\label{figflex}
\end{figure}

\begin{figure}
\centering \epsfig{width=.40\textwidth,angle=-90,file=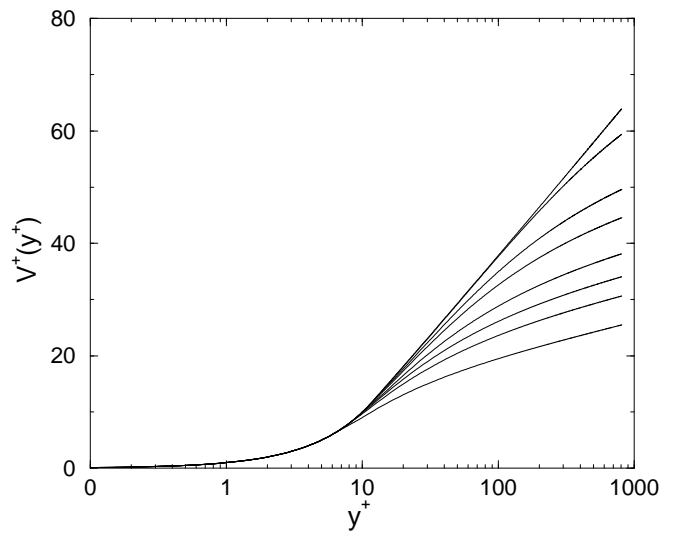}
\caption{The mean velocity profile for rodlike polymer additives
with $\tilde \nu=1,5,10,20,50, 100, 500, 1000, 5000$ and 10000 from below to above.
Note the typical behavior expected for rodlike polyemrs,  i.e. that the profile
diverges from the von K\'arm\'an log-law, reaching the MDR only asymptotically.}
\label{rod}
\end{figure}

\begin{figure}
\centering \epsfig{width=.40\textwidth,angle=-90,file=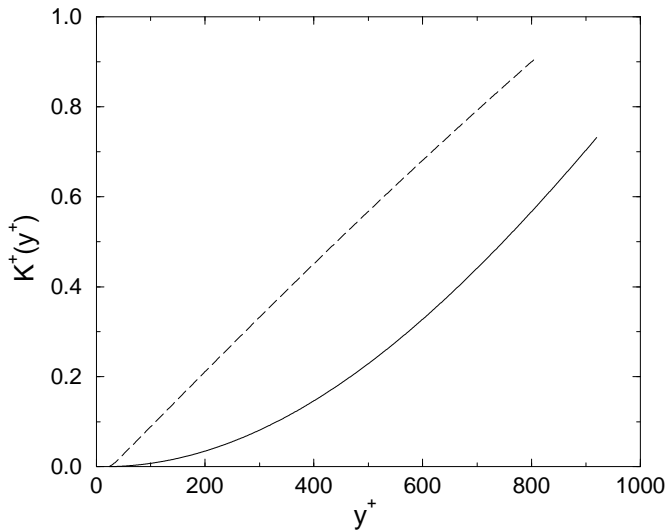}
\caption{The turbulent kinetic energy profile $K^+(y^+)$ for the flexible (solid line) and
rodlike (dashed line) cases respectively. We note the quadratic and linear dependence respectively,
as anticipated in the text.}
\label{figK}
\end{figure}

The results of the numerical solutions of the equations are shown in Figs. \ref{figflex},
\ref{rod}  and \ref{figK}.
%%%%%%% Figure 2 %%%%%%
%%%%%%% Figure 3 %%%%%%%
%%%%%%%%%%%%%%%%%%%

In Figs.~\ref{figflex} and \ref{rod} we present the mean velocity profile as a
function of the distance from the wall, for flexible and rodlike
polymers respectively. The main result of this paper is seen in
the difference between these profiles as a function of the polymer
concentration. While the flexible polymer case exhibits the feature
\cite{Virk,97VSW} that the velocity profile adheres to the MDR until a crossover
to the Newtonian plug is realized, the rodlike case presents a ``diverging"
of profiles which only asymptotically reach the MDR. We also
notice that the flexible polymer matches the MDR faithfully for
relatively low values of $\tilde \nu$, whereas the rodlike case
attains the MDR only for much higher values of $\tilde \nu$. This
result is in agreement with the experimental finding in
\cite{Bonn, Bonn2} that the flexible polymer is a better drag
reducer than the rodlike analogue.

We should note that the higher efficacy of flexible polymers
cannot be easily related to their elongational viscosity as
measured in laminar flows. In some studies \cite{Bonn, Bonn2,
Dentoonder} it was proposed that there is a correlation between
the elongational viscosity measured in laminar flows and the drag
reduction measured in turbulent flows. We find here that flexible
polymers do better in turbulent flows due to their contribution to
the effective shear viscosity, and their improved capability in
drag reduction stems simply from their ability to stretch,
something that rodlike polymers cannot do.

Finally we recall that the derivation of our equations relied on the fact that the turbulent kinetic
energy $K^+$ depends linearly and quadratically on the distance from the wall
for flexible and rodlike polymers respectively. It is important
to check that the resulting equations confirm this expectation self-consistently.
Indeed, in Fig. \ref{figK} we present the solution for this quantity in the two cases,
and find that the expectation is fully realized.
%%%%%%%% Figure 4 %%%%%%%%%%
%%%%%%%%%%%%%%%%%%%%%%%

\acknowledgments
This work has been supported in part by the European Commission under a TMR grant and by the
Research Grants Council of Hong Kong (CUHK 400304).


\begin{thebibliography}{99}
\bibitem{Book1}
A. Gyr and H. W. Bewersdorff {\em Drag Reduction of Turbulent
Flows by Additives} (Kluwe, London, 1995).

\bibitem{Virk}
P. S. Virk, AlChE Journal {\bf 21}, 625 (1975).

\bibitem{Pope}
S.B. Pope, {\em Turbulent Flows} (Cambridge, 2000).

\bibitem{Escudier}
M. P. Escudier, F. Presti and S. Smith, J. Non-Newtonian Fluid
Mech. {\bf 81}, 197 (1999).

\bibitem{97VSW}
P.S. Virk, D.C. Sherma and D.L. Wagger, AIChE Journal, {\bf 43}, 3257
(1997).

\bibitem{master}
R. Benzi, E. deAngelis, V. S. L'vov and I. Procaccia, Phys. Rev.
Lett., {\bf 95}, 194502 (2005).

\bibitem{JFM}
R. Benzi, E. de Angelis, V. S. L'vov, I. Procaccia and V. Tiberkevich,  J. Fluid
Mech. in press.

\bibitem{88DE}
M. Doi and S. F. Edwards {\em The Theory of Polymer Dynamics} (Oxford,
1988).

\bibitem{PRE}
R. Benzi, E. S.C. Ching, T. S. Lo, V. S. L'vov, and
I. Procaccia, Phys. Rev. E., {\bf 72}, 016305(2005).


\bibitem{PRL}
V. S. L'vov, A. Pomyalov, I. Procaccia and V. Tiberkervich, Phys.
Rev. Lett., {\bf 94}, 174502 (2005)

\bibitem{paperB}
V. S. L'vov, A. Pomyalov, I. Procaccia and V.
Tiberkevich, Phys. Rev. E., {\bf 71}, 016305(2005).


\bibitem{Bonn}
C. Wagner, Y. Amarouch\`ene, P. Doyle and D. Bonn, Europhys. Lett.
{\bf 64}, 823 (2003).

\bibitem{Bonn2}
D. Bonn, Y. Amarouch\`ne, C. Wagner, S. Douady and O. Cadot, J.
Phys.: Condens. Matter, {\bf 17}, S1195 (2005).

\bibitem{Dentoonder}
J. M. J. den Toonder, F. T. M. Nieuwstadt and G. D. C. KuiKen ,
Appl Sci Res,  {\bf 54}, 95 (1995).

\end{thebibliography}
\end{document}